\documentclass[showpacs,preprintnumbers,amsmath,amssymb,preprint]{revtex4}
\usepackage{graphicx}
\usepackage{dcolumn}
\usepackage{bm}
\begin{document}
\title{Temperature dependence of the Stark shifts of Er$^{3+}$ transitions in  Er$_2$O$_3$ thin film on Si (001)}

\author{Halim \surname{Choi}}
\author{Y. H. \surname{Shin}}
\author{Yongmin \surname{Kim$^{*}$}\footnotetext[1]{ Author to whom correspondence should be addressed. Electronic mail : yongmin@dankook.ac.kr }}

\affiliation{Department of Physics, Dankook University, Cheonan 31116, Korea}

\begin{abstract}
  Optical transitions of an Er$_2$O$_3$ film on a Si substrate grown by using a metal-organic chemical vapor deposition technique were investigated in a wide temperature (300 K $\sim$ 5 K) and spectral (500 nm and 850 nm) ranges.  Numerous sharp transitions corresponding to the Er$^{3+}$ ionic levels were observed, which show Stark shifts induced by the crystal field. With decreasing temperature from 300 K to 5 K, all transition peaks exhibit spectral red-shift. We believe that such red-shift behavior is due to the change of the crystal field together with the change of strain field induced by the film and the substrate in varying temperature. An interesting result is the total amount of red-shifts between 300 K and 5 K. We obtain that the higher transition energy peaks show bigger red-shifts. This is very consistent to all transition peaks. This is because the dipole moments between the transition levels are different that lead to different amounts of Stark shifts.
\end{abstract}



%\begin{keyword}
%Semiconductor, photoluminescence, diamagnetic shift
%\end{keyword}
%\end{frontmatter}
\maketitle

\section{Introduction}
Crystals containing rare earth (RE) elements are widely investigated due to their versatile usage not only for the basic researches but also for the industrial applications. Especially erbium (Er) doping in semiconductors\cite{Michel,Zavada,Bai} and glasses\cite{Burns,Chen,Feng,Baki,Huang} are of interest due to the strong emission around 1.54 $\mu$m, which can be applied to optical communication devices. The optical transitions in the trivalent Er (Er$^{3+}$) occur within the incomplete 4$\textit{f}$ shell, which is shielded by 5\textit{s} and 5\textit{p} shells.  The origin of the 1.54 $\mu$m emission is due to the recombination between $^4$I$_{9/2}$ and $^4$I$_{15/2}$ manifolds in Er$^{3+}$. When an trivalent RE element is incorporated in a crystal or forming an RE$_{2}$O$_{3}$ crystal, internal electric dipole field occurs between the RE cation and the surrounding oxide anion, which is known as the crystal field\cite{Burns2}. The crystal field generate Stark shifts in the trivalent RE ionic levels\cite{Gruber,Gruber2,Cui}. Consequently, each trivalent RE ionic levels split into many Stark-shifted energy levels. In addition to the crystal field induced Stark shifts, the exchange field\cite{Hong}, the Jan-Teller effect \cite{Lesseux} and upconversion luminescence\cite{Pavani} in Er$^{3+}$ incorporated crystals were reported.

In this study, we report temperature dependence of the optical transitions of Er$^{3+}$ levels in a Er$_{2}$O$_{3}$ film deposited on a Si (001) substrate by using a chemical vapor deposition (CVD) technique. We observed crystal field induced Stark-shifts, which show spectral red-shifts with decreasing temperature from 300 K to 5 K. Moreover, all of the transition peaks exhibit different amount of energy shift in such a way that the higher peak transition energy has bigger energy reduction with decreasing temperature. These temperature dependent optical transition behaviors are related with the thermal expansion mismatch and lattice mismatch between the film and the substrate.

\section{Experiment}

Er$_{2}$O$_{3}$  films were deposited on Si (001) substrates by using a CVD method.  Tris(2,2,6,6-tetramethyl-3,5-heptanedionato) erbium was used as an organo-metallic precursor. Four Si substrates with the size of 1 cm by 1 cm square were located at once on a substrate holder located at the reactor center. In this way, we obtained four identical Er$_{2}$O$_{3}$ film samples.  High purity Ar gas was used as a precursor carrier with the flow rate of 50 standard cubic centimeters per minute (sccm) and the flow rate of O$_{2}$ gas was 100 sccm. While growing films, the substrate temperature was 550 $^{\circ}$C and the total pressure inside of the CVD reactor was maintained at 10 mbar by using an automatic pressure regulating valve. The deposition time was 4 hours and after finishing the film deposition process, samples were slowly cooled down by 0.5 $^{\circ}$C/min.  to room temperature. Two samples were heat treated at 600 $^{\circ}$C for two hours to be compared to the remaining two as-grown samples. Specific details of the sample growth methods can be found elsewhere\cite{Choi}.

The surface morphologies, the thickness of samples and composition of the films are obtained by using SEM images and energy dispersive X-ray spectroscopy (EDS) for the as-grown and the heat treated samples, which are displayed in Fig. 1 (a) to (d). The estimated grain size of the as-grown film is ${\sim}$ 30 nm as seen in Figs. 1 (a) and (b). After the heat treatment, even though the overall grain size became larger, the effect can be ignorable. The thickness of the films were measured to be 349 nm (Fig. 1 (c)). The EDS results indicates that composition of the films are identical for both as-grown and heat treated samples(Fig. 1 (d)). For the crystallinity characteristics, an X-ray diffraction (XRD) using a Hitach Model 4600 and Raman spectroscopy were performed for both as-grown and heat treated samples. For Raman spectroscopy, a home-built micro-Raman system was used, which consists of an optical microscope equipped with an optical fiber connected to a 50-cm spectrometer containing a liquid-nitrogen-cooled charge-coupled device (CCD). The excitation laser wavelength was 532 nm from an Nd-YAG laser.

For temperature dependent photoluminescence (PL) measurements, sample temperature was varied from room temperature (300 K) to 5 K by using a closed-cycle refrigerator with optical windows. For stable temperature variation, the sample temperature was increased $\sim$ 2 min./$^{\circ}$C from 5 K to 300 K. The 442-nm line from a He-Cd  laser was used as the excitation source for the PL measurements, and a 50-cm spectrometer equipped with a liquid-nitrogen-cooled CCD was used.

\section{Results and Discussion}
	
The XRD spectra of the as-grown and heat treated films were displayed in Fig. 2 (a), which show well-defined cubic Er$_2$O$_3$ diffraction lines without any quantitative differences between two samples. Fig. 2 (b) shows the Raman spectra of the as-grown, heat treated and Si-substrate samples. The common peak at  520 cm$^{-1}$ is the characteristic peak of the Si substrate. The peaks at 800 cm$^{-1}$  and 980 cm$^{-1}$  from the Si-substrate (solid green line) maybe due to the native SiO$_2$ from the surface of the Si substrate\cite{Ivanda}. The broad peak at 800 cm$^{-1}$ disappears after the film growth, whereas the peak at 980 cm$^{-1}$ exists even after the sample growth.  It is generally known that there are four different Raman transitions within our spectral detection range, T$_g$ (580 cm$^{-1}$ and 598 cm$^{-1}$) and A$_g$ (603 cm$^{-1}$ and 622 cm$^{-1}$)\cite{Yan}. We could not obtain any of these Raman transition. In this range, the Raman peaks are closely located to the emission peaks and difficult to differentiate from the emission peaks\cite{Cui}. We concluded that even though we obtained Raman shift of Si substrate, the other peaks are related with emission peaks of $^4$S$_{3/2}$ level to $^4$I$_{15/2}$ level within the Er$^{3+}$ energy levels\cite{Yu}. The emissions from the  Er$^{3+}$ energy levels will be discussed in the following paragraph. From the results of Fig. 1 and Fig 2, we could not observe any quantitative differences between the as-grown and the heat treated samples.

Figure 3 exhibits PL spectra of the as-grown sample at 5 K (solid blue line) and room temperature (solid red line). Number indexed eleven prominent emission lines are recombination from the Er$^{3+}$ levels, which are depicted in the inset of Fig. 3. It is well known that in a solid,  the free ion energy levels of Er$^{3+}$ split into many sublevels due to the Stark effect induced by the crystal field\cite{Gruber}. Numbers 1 and 2 are identified as transitions of $^2$H$_{11/2}$, 3 from $^4$S$_{3/2}$, 4 to 10 from $^4$F$_{9/2}$  and 11 from $^4$I$_{9/2}$ to $^4$I$_{15/2}$ sublevels\cite{Chandra}. Because Er$_2$O$_3$ crystal is regarded as insulator with the energy gap of $\sim$ 5.3 eV\cite{Kamineni}, the broad background peaked at $\sim$ 1.7 eV ($\sim$ 730 nm) is due to the impurity transition\cite{Kao}. With decreasing temperature from room temperature (red line) to 5 K (blue line), the Er$^{3+}$ emission lines show spectral red-shift behavior.

One of the interesting results of the PL transitions of the Er$^{3+}$ is the energy change of the each transitions between 300 K and 5 K. Figure 4(a) displays the peak transition energy change between 300 K to 5 K for peaks number 1 to 11. With decreasing temperature, all of eleven peaks show red-shift behavior. As mentioned above, when Er$^{3+}$ ions are imbeded in a crystal, each energy levels undergo Stark shifts due to the crystal field. The transition energy change with respect to temperature is due may to the change of the crystal field caused by thermal expansion. The temperature variation of the crystal field splitting can be estimated as\cite{Burns2},
\begin{equation}
\frac{\Delta_{H}}{\Delta_{L}}=\left(\frac{V_{L}}{V_{H}}\right)^{5/3}=\left[1-\alpha(T_{H}-T_{L})\right]^{5/3},
\end{equation}
where $\Delta_{H,L}$ and $V_{H,L}$ are the crystal field splitting and the volume at high (low) temperature($T_{H,L}$), respectively, and $\alpha$ is the thermal coefficient. In this equation, because the high temperature volume is larger than that of the low temperature, the ratio becomes smaller than 1. Therefore, the low temperature crystal field splitting has to be larger than that of the high temperature. This means that with decreasing temperature, the crystal field splitting has to be increased, which is opposite to our experimental result.

To understand the reduction of the transition energy with decreasing temperature, in addition to the crystal field, one must consider  the variation of the strain field caused by the lattice mismatch at the interface between the film and the Si (001) substrate. The lattice constant of Er$_{2}$O$_{3}$ is 1.054 nm at room temperature, which is almost twice larger than that of Si (0.5431 nm)\cite{Xu}.  The  thermal expansion coefficient of Er$_{2}$O$_{3}$ at temperature between 100 K and 300 K is $\alpha_{\textrm{Er2O3}}=5.7 \times 10^{-6}$ $\textrm{K}^{-1}$\cite{Singn}, which is larger than that of Si ($\alpha_{\textrm{Si}}=2.026 \times 10^{-6}$ $\textrm{K}^{-1}$)\cite{Singn,Herrero,Dargis}. With decreasing temperature, the lattice mismatch induced tensile stress occurs because Er$_{2}$O$_{3}$ has larger lattice constant and the expansion coefficient than those of Si substrate. The lattice mismatch induced stress change to the film has to be considered with varying temperature. Therefore, the large lattice constant and thermal expansion mismatch induced strain field dominates the crystal field, which leads to the spectral red-shift in the optical transition in Er$_{2}$O$_{3}$ film with decreasing temperature.

%lowering temperature changes the strain field decreases the crystal field splitting energy. Therefore, these two sources of the internal electric field, which split the transition peaks compete with decreasing temperature. Because

%The pressure dependent crystal field splitting can be estimated as\cite{Burns},
%\begin{equation}
%\frac{\Delta_{p}}{\Delta_{0}}=\left(\frac{V_{0}}{V_{p}}\right)^{-5/3},
%\end{equation}
%where $\Delta_{p,0}$ and $V_{p,0}$ are the crystal field splitting and the volume at high (ambient) pressure, respectively. According to Eq. 2, increasing pressure increases the crystal field splitting energy. Because the lattice mismatch induced stress at the interface is not uniaxial and the Er$_{2}$O$_{3}$ film used for this study is not a single crystal, this equation cannot be applied  directly to our case. However, the effect is limited, it is obvious that the relaxed strain field while decreasing temperature at the interface reduces the pressure to the film which leads to decrease the crystal field splitting energy. 

The total amount of the energy shift, which is the shift between 5K and 300 K is displayed in Fig.4 (b). It is clearly shown that the higher energy peaks show bigger energy reduction in a consistent way. The total energy shift of peak 1 ($^2$H$_{11/2}$) exhibit 2.9 meV, whereas peak 11 ($^4$I$_{9/2}$) shows only 1 meV shift. It can be explained in such a way that the amounts of the Stark shift of different energy levels have to be different because different levels have different dipole moments. In this regard, each transitions show different energy shifts with varying temperature.

%As temperature decreased, due to the thermal shrinkage of the crystal, not only the crystal field but also the strain field changes due to the different thermal expansion of the film and the substrate. Therefore, the variation of the transition energy is the result caused by the change of the crystal field and the interface strain field.

\section{Conclusion}

We deposited an Er$_{2}$O$_{3}$ film on a Si substrate by using a CVD method. The thickness of the film was measured to be 395 nm by using an SEM image. The surface images taken by an SEM show that grains with average size of $\sim$ 30 nm for an as-grown sample. X-ray diffraction and Raman analysis indicate that there is no quantitive difference between the as-grown and the heat treated samples. PL measurements exhibit sharp emission lines, which are transitions within the Er$^{3+}$ ionic energy levels. Multiple-transition occurs within a manifold due to the internal field induced Stark shift. The temperature dependent PL transition measurements suggest that in addition to the crystal field, one must consider the interface strain field between the Er$_2$O$_3$ and the Si substrate. With decreasing temperature, due to the large thermal expansion and the lattice mismatches, all transition peaks undergo red-shifts. Lowering temperature has to increase the transition energy, which is opposite to the experimental result. Therefore, the thermal expansion and lattice mismatch induced strain field dominates the crystal field effect, which leads to spectral red-shift with decreasing temperature. The total amounts of the red-shifts for each transitions are different in such a way that the higher transition energy peaks undergo larger red-shift. This is because the dipole moments within the Er$^{3+}$ manifolds are different.

\section*{Acknowledgments}
The present research was supported by the research fund of Dankook University in 2016.

\section*{References}
%\bibliographystyle{model1a-num-names}

%\begin{thebibliography}{00}

%\end{thebibliography}

\newpage

\begin{figure}
\caption{Fig 1 (a) and (b) Scanning Electron Microscope  images of as-grown and heat treated samples, respectively. Both samples show granular-shape polycrystals with the average grain size of $\sim$ 30 nm. (c) The thickness of the Er$_{2}$O$_{3}$ film is 349 nm. (d) EDS spectra of the as-grown (dashed red line) and the heat treated (solid black line) samples. Both samples exhibit the identical amounts of each contents of elements.}
\label{fig1}
\end{figure}

\begin{figure}
\caption{Fig. 2 (a) X-ray diffraction (XRD) peaks of  as-grown and heat treated Er$_{2}$O$_{3}$ film on Si substrates. Both samples show identical XRD peaks. (b) Raman spectra of as-grown (blue line), heat treated (red-line) Er$_{2}$O$_{3}$ films and Si substrate (green line). }
\label{fig2}
\end{figure}

\begin{figure}
\caption{Fig. 3 Photoluminescence transition spectra of as-grown sample at 300 K (red line) and at 5 K. Eleven prominent peaks are observed, which are transitions within Er$^{3+}$ levels. The inset shows schematic diagram of possible transitions. Multiple transition in a manifold can be possible due to the Stark shift induced by the crystal field.}
\label{fig3}
\end{figure}

\begin{figure}
\caption{Fig. 4 (a) The temperature dependence of the peak transition energy between 300 K and 5 K. All transitions show spectral red-shift. (b) The total amount of energy shift between 300 K and 5 K. The total energy shift consistently show in such a way that the higher transition energy exhibits bigger energy shift.}
\label{fig5}
\end{figure}

\end{document}